# Energy-Delay Efficient Power Control in Wireless Networks

Alessio Zappone, *Senior Member, IEEE*, Luca Sanguinetti, *Senior Member, IEEE*, Mérouane Debbah, *Fellow, IEEE*

*Abstract*—This work aims at developing a power control framework to jointly optimize energy efficiency (measured in bit/Joule) and delay in wireless networks. A multi-objective approach is taken to deal with both performance metrics, while ensuring a minimum quality-of-service to each user in the network. Each user in the network is modeled as a rational agent that engages in a generalized non-cooperative game. Feasibility conditions are derived for the existence of each player's best response, and used to show that if these conditions are met, the game best response dynamics will converge to a unique Nash equilibrium. Based on these results, a convergent power control algorithm is derived, which can be implemented in a fully decentralized fashion. Next, a centralized power control algorithm is proposed, which also serves as a benchmark for the proposed decentralized solution. Due to the non-convexity of the centralized problem, the tool of maximum block improvement is used, to trade-off complexity with optimality.

## I. INTRODUCTION

The number of wirelessly connected devices is growing at an exponential rate, and is anticipated to reach 50 billions by 2020 [1]. This poses serious concerns for the design of the fifth generation (5G) of cellular networks. In order to serve so many devices, 5G networks will have to provide $1000\times$ higher data rates compared to present networks [2]. However, achieving this by simply scaling up the transmit power is not possible, because it would result in an unmanageable energy demand, besides causing alarming greenhouse gas emissions and electromagnetic pollution [3]. It is common belief that the $1000\times$ data rate increase must be achieved while, at the same time, halving the power consumption [3]. This requires a $2000\times$ increase of the bit-per-Joule energy efficiency in 5G networks compared to present systems. A recent collection of advanced energy-efficient solutions for 5G can be found in [4]. Moreover, the exponential growth of wirelessly connected devices makes it very difficult to perform traditional centralized network designs, which would lead to huge feedback overheads and prohibitive computational complexity. This calls for distributed network designs, which allow

A. Zappone, L. Sanguinetti, and M. Debbah are with the Large Systems and Networks Group (LANEAS), CentraleSupélec, Université Paris-Saclay, 3 rue Joliot-Curie, 91192 Gif-sur-Yvette, France. A. Zappone is also with the University of Cassino and Southern Lazio, Cassino, Italy (alessio.zappone@unicas.it). L. Sanguinetti is also with the University of Pisa, Dipartimento di Ingegneria dell'Informazione, Pisa, Italy (luca.sanguinetti@iet.unipi.it). M. Debbah is also with the Mathematical and Algorithmic Sciences Lab, Huawei France R&D, Paris, France (merouane.debbah@huawei.com)

The research of A. Zappone by the H2020 MSCA IF BESMART, Grant 749336. The work of M. Debbah and L. Sanguinetti was supported by the ERC Starting Grant 305123 MORE.

A preliminary version of this work has been presented at the 2016 Workshop on Smart Antennas (WSA 2016), Munich, Germany.

for self-organizing networks that drastically reduce feedback transmissions and computational complexity. A recent survey on self-organizing networks for 5G is [5]. Another important requirement for 5G networks is to reduce the communication latency down to $1\,\mathrm{ms}$, a requirement which is not met by present LTE systems [6]. The $1\,\mathrm{ms}$ constraint corresponds to the human response time to visual feedback control. Therefore, a communication delay inferior to this threshold would enable the so-called tactile internet concept [7], i.e. the use of wireless communications as a platform to control and steer real and virtual objects in many practical situations, with applications to health-care, mobility, education, manufacturing, smart grids, just to name a few. Achieving this goal appears quite a challenging task, especially in light of the huge size of future 5G networks.

### A. Literature review

The above considerations motivate the need for the investigation of energy-efficient, distributed, and delay-aware power allocation algorithms in wireless networks. However, the majority of available contributions consider energy-efficient and distributed schemes, without also including the delay aspect in the algorithm design. The energy efficiency of each user is defined as the ratio between the user's throughput and consumed power, and the tools of game theory are used to model the interactions among the different network nodes [8], [9]. More specifically, the network user terminals are modeled as utility-driven, rational agents that engage in a non-cooperative game for individual energy efficiency maximization [10], [11]. In [12], game theory is used to perform energy-efficient transceiver design in multiple-access networks, whereas in [13] the results from [12] are extended to systems using widely linear processing. In [14], the Nash equilibrium (NE) problem for a group of players aiming at maximizing their own energy efficiency (EE) while satisfying power constraints in single and multicarrier systems is studied. A quasi-variational inequality approach is taken in [15], where power control algorithms for networks with heterogenous users are developed. In [16], [17], a similar problem is considered for relay-assisted systems. A common drawback of all of these works is that no rate requirement is taken into account. This might lead to operating points where some terminals experience fairly low rates. Nevertheless, imposing individual target rates changes the setting drastically since any user's admissible power allocation policy depends on the transmit policies of all other users. This problem has been studied in [18] wherein Nash equilibria are found to be the fixed

4points of a water-filling best-response operator whose water level depends on the rate constraints and circuit power, and in [19] wherein a general framework for non-cooperative EE maximization is provided, encompassing several 5G candidate technologies.

As already mentioned, all previous works do not consider the communication delay in the resource allocation process. Indeed, very few works have considered a holistic approach considering distributed designs which are at the same time energy-efficient and delay-aware. The energy-delay tradeoff in single-user systems is discussed in [20]. A multiuser network is considered in [21], and the communication delays are included as constraints in the resource allocation process, deriving non-cooperative energy efficiency maximization algorithms subject to minimum delay guarantees for the individual communication links. However, this approach does not allow the optimization of the network resources to minimize the communication delays. A more recent approach based on large deviation theory embeds the communication delay into the performance function by introducing a delay parameter into the classical Shannon capacity formula, obtaining the so-called effective capacity function [22]. Then, an energy-efficient and delay-aware performance function is defined by the ratio between the effective capacity and the consumed power, a metric called effective energy efficiency [23], [24]. However, the main goal of this method is on stabilizing the queue of the transmit buffer, rather than minimizing the transmission delay. Instead, an approach that allows the simultaneous optimization of both energy efficiency and communication delay has been proposed in [25]. Therein, a multi-objective approach is taken, where a single objective function is formulated based on both energy and delay costs of communication. Then, based on these individual energy-efficient and delay-aware performance functions, a non-cooperative game is formulated, showing that it admits a unique Nash equilibrium, which can be reached by implementing the game best-response dynamics.

*B. Contributions and outline*

This work extends the multi-objective optimization approach from [25] in several directions and provides the following major contributions.

• The framework proposed in [25] is extended to include quality-of-service (QoS) constraints in terms of maximum bit error rate or minimum achievable rate. This makes the problem much more challenging, because the non-cooperative game to be analyzed becomes a generalized game, in which not only the users' utility functions, but also the users' strategy sets are coupled. Moreover, a more general users' signal to interference plus noise ratio (SINR) expression is considered compared to [25]. In particular, the analysis is performed assuming that the users' SINRs follow the model from [19]. This makes the energy-efficient and delay-aware framework proposed in this work general enough to encompass several leading technologies for 5G networks, such as Massive multiple-input multiple-output (MIMO), heterogeneous networks, and full-duplex communications.

• Within this challenging scenario, first closed-form feasibility conditions are derived for the individual best-response problems of the considered generalized non-cooperative game. Next, it is shown that, when these conditions are met, the generalized non-cooperative game admits a unique Nash equilibrium which can be reached by implementing the game best-response dynamics. These results enable to devise a fully distributed, energy-efficient, and delay-aware power control algorithm.

• A centralized power control algorithm is derived, which also serves as a benchmark for the distributed solution. The non-convex centralized power control problem is tackled by the method of maximum block improvement [26], which is guaranteed to converge to good candidate solutions with affordable complexity.

• The energy and delay performance of the proposed methods are numerically evaluated. Among the several applications of the framework, a Massive MIMO system is selected, considering the effect of imperfect channel state information (CSI) at the receiver and the presence of hardware impairments.

The remainder of this paper is organized as follows. Section II introduces the signal model and formulates the distributed power control problem from a game-theoretic perspective. Section III carries out the analysis of the non-cooperative game and develops the corresponding distributed power control algorithm. Section IV describes a centralized algorithm, which can be used for benchmarking purposes. The performance of the proposed algorithms are numerically evaluated in Section V, considering a Massive MIMO setup. Concluding remarks are given in Section VI. The basic results from generalized convexity and fractional programming as well as the theoretical setup of the centralized optimization framework are revised in the Appendices.

## II. SYSTEM MODEL

Consider the uplink of a wireless interference network[1], with $K$ transmitters and $M$ receivers and let the SINR of user equipment (UE) $k$ take the following general form:

$$\gamma_k = \frac{p_k \alpha_k}{\sigma_k^2 + \phi_k p_k + \sum_{j \neq k} p_j \beta_{k,j}}. \quad (1)$$

In (1), $p_k$ is the transmit power of user $k$, $\{\sigma_k^2\}_k$, $\{\alpha_k\}_k$, $\{\phi_k\}_k$, and $\{\beta_{k,j}\}_{k,j}$ are coefficients which fulfill the following three assumptions, for any $k$ and $j$:

• they are non-negative real numbers, which depend on system parameters, channel gains, and thermal noise, but not on the users' transmit powers;

• $\sigma_k^2$ models the thermal noise at the receiver associated to user $k$, but can also depend on user $k$'s channel and on global system parameters;

• $\alpha_k$ and $\phi_k$ only depend on user $k$'s channel and possibly on global system parameters, while $\{\beta_{j,k}\}_j$ depend on the channel from transmitter $k$ to receiver $j$ and possibly on global system parameters.

The explicit expressions for $\{\sigma_k^2\}_k$, $\{\alpha_k\}_k$, $\{\phi_k\}_k$, and $\{\beta_{j,k}\}_{j,k}$ depend on the specific network under analysis, and 3we hasten to stress that many relevant instances of communication systems can be modeled by (1), by suitably specifying $\{\sigma_k^2\}_k$, $\{\alpha_k\}_k$, $\{\phi_k\}_k$, and $\{\beta_{j,k}\}_{j,k}$. Besides the simpler case in which $\phi_k = 0$ for all $k$, leading to the familiar SINR expression encountered in wireless communication systems, the presence of non-zero $\{\phi_k\}_k$ allows modeling several 5G candidate technologies. Examples in this sense include: practical massive MIMO networks subject to hardware impairments and/or imperfect channel estimation at the receiver [19]; relay-assisted networks [16], [27]; device-to-device (D2D) communications [28]. A detailed description of these case-studies is reported in [16], [19], [27], together with the corresponding expressions of the coefficients in (1). In addition, other notable communication scenarios which lead to the SINR form in (1) are systems affected by inter-symbol interference and/or frequency-selective fading [29], [30].

---

[1]The mathematical results apply to any interference network. Nevertheless, the implementation of the distributed resource allocation algorithm to be developed fits better an uplink scenario.

*A. Metrics*

In this work, we are interested in two relevant performance metrics for a communication link, namely, the transmission delay and the energy consumption. As for the transmission delay, following [25] we consider a system in which at each time-slot a packet arrives at the transmitter queue of UE $k$ with probability $\lambda_k$, and assume that packet arrival events are statistically independent of each other as well as of transmission success and failure events. Under these assumptions, the average time required for a reliable transmission can be expressed as [25]:

$$D_k = \frac{1}{R}\frac{1}{S_k(\gamma_k) - \lambda_k} \quad (2)$$

where $S_k(\gamma_k)$ is the probability of correct packet reception and $R$ is the data communication rate measured in [bit/s] and assumed, without loss of generality, to be the same over all links.[2] The metric in (2) is measured in [s/bit] and represents the amount of time per reliably transmitted bit over link $k$. Notice that (2) represents a valid delay only if $S_k(\gamma_k) - \lambda_k > 0$. From a physical point of view, this means that the amount of data arriving at UE $k$ must be lower than the amount of data that UE $k$ can reliably transmit per unit of time – otherwise, the transmit buffer would overflow, thereby causing data loss.

To capture the trade-off between reducing energy consumption and guaranteeing reliable data communications, we consider the cost-benefit ratio of the link expressed as the ratio between the consumed power and the corresponding amount of data reliably received. This leads to:

$$F_k = \frac{p_k + P_{C,k}}{RS_k(\gamma_k)} \quad (3)$$

wherein $P_{C,k}$ is the static hardware power dissipated in all circuit blocks, except the transmit amplifier, required to operate link $k$.[3] The measure unit of (3) is [Joule/bit] as it represents the amount of energy required to transmit a given amount of data or, otherwise stated, the energy cost per reliably transmitted bit.[4]

The probability of correct packet reception $S_k$ depends on the specific communication system and it can be a very involved function (or even not available in closed-form). A widely used approximation for $S_k(\gamma_k)$ is [16], [25]:

$$S_k(\gamma_k) = 1 - e^{-\delta_k \gamma_k} \quad (4)$$

where $\delta_k > 0$ depends (among other system parameters) on the modulation scheme used by user $k$ and the radio propagation environment, and on the packet length. The following analysis is, however, not limited to a particular expression of $S_k(\gamma_k)$ but applies to any function $S_k(\gamma_k)$ with the following properties:

1) $S_k(\gamma_k) \geq 0\ \forall \gamma_k \geq 0$ with $S_k(0) = 0$, i.e. a non-negative amount of data is transmitted $\forall \gamma_k \geq 0$, but no data is sent if no transmit power is used. In this latter case, the energy cost (3) tends to infinity.
2) $S_k(\gamma_k)/\gamma_k \to 0$ for $\gamma_k \to \infty$, i.e. by using an infinite amount of power, the energy cost diverges.
3) $S_k(\gamma_k)$ is increasing $\forall \gamma_k \geq 0$, i.e. more data can be sent by spending more power.
4) $S_k(\gamma_k)$ is concave $\forall \gamma_k \geq 0$.

*Remark 1:* It is easy to check that (4) fulfills all the above properties. Moreover, another relevant performance function, which fulfills all the above properties, is the channel achievable rate $W \log_2(1 + \gamma_k)$, with $W$ being the communication bandwidth. This latter choice is also very popular [14], [32], and does not affect the measure unit of (2) and (3). However, it should be emphasized that while the denominator of (3) accounts for the actual communication goodput, i.e. the amount of bits that are reliably transmitted per unit of time, the achievable rate is an upper-bound to the actual goodput, namely measuring the amount of bits that *can be reliably transmitted* per unit of time.

*Remark 2:* While Properties 1-3 stem from physical considerations (as explained above), Property 4 does not necessarily apply to all physically meaningful functions $S_k(\gamma_k)$. For example, it is not satisfied by the following function (used in several previous works [12], [30])

$$S_k(\gamma_k) = \left(1 - e^{-\gamma_k}\right)^Q \quad (5)$$

with $Q$ being the packet length. It is seen that (5) is not concave in $\gamma_k$, and thus it is not included in the framework developed henceforth. Nevertheless, it should be mentioned that the two forms in (4) and (5) are closely related, being two different approximations of the probability of packet reception [12], [25]. The expression in (5) approximates by $1 - e^{-\gamma_k}$ the probability of correct reception of a single bit in the packet. By assuming independent symbol transmissions, (5) is thus obtained. Instead, (4) directly approximates the probability of correctly receiving the whole packet. To this end, standard numerical techniques can be used to determine the parameter $\delta_k > 0$ so as to improve the accuracy of the approximation.

---

[2] A user-dependent communication rate $R_k$ could easily be handled by defining $\bar{S}_k(\gamma_k) = R_k S_k(\gamma_k)$.

[3] Note that the efficiency $\eta_k$ of the linear power amplifier at the transmitter can be easily accounted for scaling $p_k$ by $\mu_k = 1/\eta_k \geq 1$.

[4] The quantity in (3) can be seen to be the inverse of the so-called energy efficiency of link $k$, which is a more widely used, yet equivalent, metric to measure the efficiency with which energy is used to transmit data [31].

To find a balance between the two utility functions in (2) and (3) over link $k$, we apply the *scalarization* technique [33], and define the overall cost function over link $k$ as:

$$c_k = \rho_k D_k + F_k = \frac{1}{R}\left(\frac{\rho_k}{S_k(\gamma_k) - \lambda_k} + \frac{p_k + P_{C,k}}{S_k(\gamma_k)}\right) \quad (6)$$

wherein $\rho_k > 0$ is a positive coefficient measured in [Joule/s], which weighs the relative importance of $D_k$ with respect to $F_k$. We stress that the two cost functions in (2) and (3) can in principle be combined in many other ways. Specifically, the theory of multi-objective optimization ensures that the optimization of any combining function $g(D_k, F_k)$, decreasing in both arguments, yields a Pareto-optimal point in the space of all feasible pairs $(F_k, D_k)$, [34]–[36]. Here, the focus is on a linear combining function $g$, leaving for future works the extension to a generic one. Based on (6), the powers $\{p_k; k = 1, \ldots, K\}$ can be optimized following a distributed or a centralized paradigm. Both approaches will be considered in the following. The distributed solution is illustrated first.

## III. Distributed power control

### A. Game Formulation

In a distributed approach, each UE $k$ aims at locally minimizing its own cost function $c_k$. This can be mathematically modeled by considering UEs as independent decision-makers, which engage in a non-cooperative game given by (in normal form) [9]:

$$\mathcal{G} = \{\mathcal{K}, \{\mathcal{A}_k\}_{k=1}^K, \{c_k\}_{k=1}^K(p_k, \mathbf{p}_{-k})\} \quad (7)$$

wherein $\mathcal{K} = \{1, \ldots, K\}$ is the players' set, $\mathbf{p}_{-k} = [p_1, \ldots, p_{k-1}, p_{k+1}, \ldots, p_K]$, while $\mathcal{A}_k$ is the action set of player $k$ that basically defines the feasible set of $p_k$. We take

$$\mathcal{A}_k = \{p_k \in \mathbb{R} : 0 \leq p_k \leq P_{\max,k},\ S_k(\gamma_k) \geq \theta_k\} \quad (8)$$

where $P_{\max,k}$ denotes the maximum transmit power and $\theta_k$ accounts for the minimum QoS requirement over link $k$. We assume $\theta_k > \lambda_k$, recalling that the SINR-range of interest is $\gamma_k > S^{-1}(\lambda_k)$. Given the above notation, the best response (BR) of player $k$ to a given power vector $\mathbf{p}_{-k}$ (chosen by the other players) is determined as the solution to [9]:

$$\arg\min_{p_k} \quad c_k(p_k, \mathbf{p}_{-k}) \quad (9a)$$
$$\text{subject to} \quad p_k \in \mathcal{A}_k \quad (9b)$$

where we have explicitly shown the functional dependence of $c_k$ from other players' powers $\mathbf{p}_{-k}$. The coupled problems (9) for $k = 1, \ldots, K$ define the best response dynamics (BRD) of $\mathcal{G}$, and any fixed point, if any, of the BRD is an NE of $\mathcal{G}$. The main challenges posed by $\mathcal{G}$ in (7) can be summarized as follows. Firstly, both the cost function $c_k$ and the action set $\mathcal{A}_k$ of player $k$ are coupled in $\mathcal{G}$, since $\mathcal{A}_k$ depends on the SINR $\gamma_k$ which is a function of the other players' powers $\mathbf{p}_{-k}$. A non-cooperative game of this form is known as a *generalized* non-cooperative game [37], [38], whose analysis is typically more involved than non-cooperative games. Secondly, the cost functions $\{c_k; k = 1 \ldots, K\}$ are not expressed as a single fraction given by the ratio of a convex over a concave function. This prevents from immediately concluding that $c_k$ is quasi-convex, which is one of the required conditions for the existence of an NE in *generalized* non-cooperative games. Third, $\gamma_k$ is a fractional function of $p_k$ due to the presence of the non-zero coefficient $\phi_k$. This complicates further the analysis of $\mathcal{G}$ compared to the canonical case in which $\gamma_k$ is linear in $p_k$. In what follows, we first study the feasibility of the generic BR problem (9), then provide sufficient conditions such that a unique NE exists and the BRD converges to such equilibrium

### B. Feasibility

For notational convenience, let $\omega_k = \sigma_k^2 + \sum_{j \neq k} p_j \beta_{k,j}$ so that $\gamma_k$ in (1) can be rewritten as

$$\gamma_k = \frac{p_k \alpha_k}{\phi_k p_k + \omega_k}. \quad (10)$$

*Lemma 1:* A sufficient condition for (9) to be feasible for any $\mathbf{p}_{-k}$ is the pair of inequalities:

$$S_k\left(\frac{\alpha_k}{\phi_k}\right) > \theta_k \quad (11a)$$

$$P_{\max,k} \geq \frac{S_k^{-1}(\theta_k)\left(\sigma_k^2 + \sum_{j \neq k} \beta_{k,j} P_{\max,j}\right)}{\alpha_k - S_k^{-1}(\theta_k)\phi_k}. \quad (11b)$$

*Proof:* Condition (11a) follows by the observation that, given (1) and Property 3, $S_k(\gamma_k) \leq S_k(\alpha_k/\phi_k)$, for all $p_k \geq 0$. As for (11b), in order to meet $S_k(\gamma_k) \geq \theta_k$, $p_k$ must be such that

$$p_k \geq \frac{S_k^{-1}(\theta_k)\omega_k}{\alpha_k - S_k^{-1}(\theta_k)\phi_k}. \quad (12)$$

The right-hand side (RHS) of (12) is guaranteed to be positive due to (11a). In order for (12) to hold for any $\mathbf{p}_{-k}$, the worst-case scenario in which the RHS is maximized with respect to $\mathbf{p}_{-k}$ must be considered. This is obtained by assuming $p_j = P_{\max,j}$ for all $j \neq k$, which yields (11b). ∎

It is interesting to observe that when $\phi_k \to 0$ (11a) is always verified. Indeed, if no self-interference is suffered by user $k$, it holds $S_k(\gamma_k) \to \infty$ when $p_k \to \infty$. Thus, in this case the value of $S_k(\gamma_k)$ is only limited by the value of $P_{\max,k}$.

It is worth stressing that the result in Lemma 1 provides a sufficient condition ensuring that all best-response problems in the BRD are feasible. Thus, the condition in Lemma 1 allows checking the feasibility of the complete BRD in an off-line manner, before the BRD has even started. On the other hand, a necessary and sufficient condition for the feasibility of the best-response problem (9) is obtained recalling that the minimum power required to meet (11b) is given by (12), which is a positive quantity provided that (11a) holds. Thus, (9) is feasible if and only if (11a) is true and

$$P_{\max,k} \geq \frac{S_k^{-1}(\theta_k)\omega_k}{\alpha_k - S_k^{-1}(\theta_k)\phi_k}\ . \quad (13)$$

Nevertheless, for each $k$, (13) depends on the other users' transmit powers via the parameter $\omega_k$, and thus does not lend itself to being checked off-line, but must be verified online, before solving each best-response of the BRD.

On the other hand, an off-line necessary condition for the feasibility of the complete BRD is obtained by considering the feasibility test

$$\max_{\{p_k\}_{k=1}^K} 1 \tag{14a}$$

$$\text{subject to} \quad 0 \leq p_k \leq P_{\max,k}, \ \forall k \tag{14b}$$

$$S(\gamma_k) \geq \theta_k, \ \forall k. \tag{14c}$$

Clearly, if (14) is not feasible, then, for any $k$, Problem (9a) will be unfeasible regardless of the other users' transmit powers $\{p_j\}_{j \neq k}$. Paralleling the approach from [19, Lemma 1], it can be shown that (14) is feasible if and only if it holds

$$\rho_{\mathbf{F}} < 1 \quad \text{and} \quad (\mathbf{I} - \mathbf{F})^{-1} \mathbf{S} \leq \mathbf{P}_{\max}, \tag{15}$$

where $\mathbf{S} = \left\{ \frac{\sigma_k^2 S_k^{-1}(\theta_k)}{\alpha_k - \phi_k S_k^{-1}(\theta_k)} \right\}_{k=1}^K$, $\mathbf{P}_{\max} = \{P_{\max,k}\}_{k=1}^K$, $\mathbf{F}$ is the matrix

$$[\mathbf{F}]_{k,j} \triangleq \begin{cases} 0 & j = k \\ \frac{\beta_{j,k} S_k^{-1}(\theta_k)}{\alpha_k - \phi_k S_k^{-1}(\theta_k)} & j \neq k \end{cases} \tag{16}$$

and $\rho_{\mathbf{F}}$ its spectral radius. Thus, (15) is a necessary feasibility condition for the game BRD, which can be checked off-line, before starting the BRD, since it does not depend on any transmit power.

### C. Analysis of $\mathcal{G}$ and convergence of BRD

Having determined a condition such that the BR problem (9) is feasible, the goal of this section is to: *i)* solve (9); *ii)* determine if $\mathcal{G}$ admits one or more NE; and *iii)* understand if the BRD of $\mathcal{G}$ is guaranteed to converge to an NE from any initialization point. To begin with, we provide the following result.

*Proposition 1:* If (9) is feasible, then its solution $p_k^\star$ is given by

$$p_k^\star = \min\{P_{\max}, \max\{P_{\min,k}, \bar{p}_k\}\} \tag{17}$$

where

$$P_{\min,k} = \frac{S_k^{-1}(\theta_k)\omega_k}{\alpha_k - S_k^{-1}(\theta_k)\phi_k} \tag{18}$$

and $\bar{p}_k$ is the unique stationary point of $c_k$, i.e., $\frac{\partial c_k(\bar{p}_k)}{\partial p_k} = 0$.

*Proof:* The result follows from the first-order optimality conditions of (9). The details of the proof are reported in Appendix A. ∎

An immediate consequence of Proposition 1 is the following corollary.

*Corollary 1:* If (9) is feasible for any $k$, then $\mathcal{G}$ admits an NE.

*Proof:* A generalized game admits an NE provided that [37]:

1) The players' feasible action sets $\mathcal{A}_k(\mathbf{p}_{-k})$ are nonempty, closed, convex, and contained in some compact set $\mathcal{C}_k$ for all $\mathbf{p}_{-k} \in \mathcal{A}_{-k} \equiv \prod_{\ell \neq k} \mathcal{A}_\ell$.
2) The sets $\mathcal{A}_k(\mathbf{p}_{-k})$ vary continuously with $\mathbf{p}_{-k}$, i.e. the graph of the set-valued correspondence $\mathbf{p}_{-k} \mapsto \mathcal{A}_k(\mathbf{p}_{-k})$ is closed.
3) Each user's cost function $c_k(p_k, \mathbf{p}_{-k})$ is quasi-convex in $p_k$ for all $\mathbf{p}_{-k} \in \mathcal{A}_{-k}$.

For the case at hand, the sets $\mathcal{A}_k(\mathbf{p}_{-k})$ are nonempty by assumption. Moreover, they are closed and bounded for every $\mathbf{p}_{-k}$, as well as convex, since the constraints, which define $\mathcal{A}_k(\mathbf{p}_{-k})$, are convex. Moreover, each of them varies continuously with $\mathbf{p}_{-k}$ since $S(\gamma_k) \geq \theta_k$ is itself continuous in $\mathbf{p}_{-k}$. Finally, the quasi-convexity of $c_k$ directly follows from the proof of Proposition 1, and in particular from the fact that the derivative of $c_k$ has a unique zero $\bar{p}_k$, being negative for $p_{\lambda_k} < p_k \leq \bar{p}_k$ and positive for $p_k > \bar{p}_k$, with $p_{\lambda_k}$ the power level such that $S(\gamma_k) = \lambda_k$. ∎

The above results show that, under feasibility conditions, $\mathcal{G}$ admits at least one NE. The next step is to understand how many NE exist, and if the BRD dynamics of $\mathcal{G}$ is guaranteed to converge to an NE. This is particularly important because, if multiple equilibria exist, then the issue arises of understanding how to select the most efficient NE. Also, having a convergent BRD constitutes the basis for the development of a distributed algorithm to compute an NE. To address this important point, we start with showing the following lemma.

*Lemma 2:* If

$$S_k(\gamma_k) S_k'(\gamma_k) - \gamma_k (S_k'(\gamma_k))^2 + \gamma_k S_k(\gamma_k) S_k''(\gamma_k) \leq 0 \tag{19}$$

holds for any $\gamma_k \geq 0$ we have that

$$\frac{\partial^2 c_k}{\partial p_k \partial \omega_k} \leq 0, \ \forall \omega_k > 0. \tag{20}$$

*Proof:* The proof is given in Appendix B. ∎

Using Lemma 2, it can be established that $\mathcal{G}$ admits a unique NE and that its BRD always converges to this unique NE.

*Proposition 2:* If (9) is feasible for all $k$ and (19) holds, then $\mathcal{G}$ in (7) admits a unique NE, and the BRD is guaranteed to converge to the unique NE.

*Proof:* Under the feasibility conditions for (9) given in Lemma 1, we have already shown in Proposition 1 that (7) admits an NE. Then, the NE is unique and the BRD converges to such NE provided that the BR function (17) is a standard function, as defined in [39]. This can be shown by using the results of Lemma 2 following similar arguments as those in [19, Proposition 4]. ∎

### D. Distributed implementation

Being ensured that the game BRD is globally convergent, the unique NE of the game can be reached by implementing the game BRD in (17). The critical step towards the distributed implementation of the BRD (17) is that the parameter $\omega_k$ depends on the other players' powers and channels, and therefore it is not locally available to player $k$. However, this issue can be overcome by noticing that $\omega_k$ can also be expressed as a function of $\gamma_k$ as

$$\omega_k = \frac{\alpha_k p_k}{\gamma_k} - \phi_k p_k. \tag{21}$$

The advantage of this reformulation is that $\gamma_k$ is locally available for link $k$, because it can be measured at the receiver associated to UE $k$, and fed back by a return downlink channel which is typically available in most wireless communication systems. We stress that such an approach does not require any overhead communication between a given receiver and the UEs associated to different receivers, but only between a receiver and its associated UEs. Finally, as for the parameters $\alpha_k$ and $\phi_k$, they can be locally computed as they only depend on the channel coefficient of UE $k$. Regarding this point it must be stressed that (21) assumes perfect knowledge of both $\alpha_k$ and $\phi_k$, which is in line with the perfect CSI scenario considered in this work. Nevertheless, in case of non-negligible parameter estimation errors, it is possible to either compute (21) based on the estimated $\alpha_k, \phi_k$, or to employ robust resource allocations, considering the worst-case estimation error[5] [40]–[42].

Based on (21), a distributed implementation of the game BRD can be formulated as in Algorithm 1, which converges to the unique NE of $\mathcal{G}$ by virtue of Proposition 2. An important point to stress about Algorithm 1 is that, although it is stated here assuming synchronous power updates among the users, convergence to a generalized Nash equilibrium holds also in case of asynchronous updates. Indeed, the proof that the best-response correspondence of $\mathcal{G}$ is a standard function does not depend on the way the powers are updated, but only on the mathematical properties of the best-response problems. Then, the standard property ensures the convergence also with asynchronous updates [39].

Finally, if the sufficient feasibility condition derived in Lemma 1 holds, a feasible initialization point for Algorithm 1 always exists. For example, a feasible power vector is $p_k = P_{\max,k}$ for all $k = 1, \ldots, K$. Alternatively, if (15) holds, a feasible initialization power vector can be found by solving the feasibility test (14).

---

**Algorithm 1** Distributed Power Control

**Initialize** $p_k$ to feasible values for $k = 1, \ldots, K$;
**Compute** $\alpha_k$ and $\phi_k$ for $k = 1, \ldots, K$; **Set** $\varepsilon > 0$
**repeat**
  **for** $k = 1$ to $K$ **do**
    $\omega_k = \frac{\alpha_k p_k}{\gamma_k} - \phi_k p_k$;
    $p_k = \min\{P_{\max}, \max\{P_{\min,k}, \bar{p}_k\}\}$;
  **end for**
**until** Convergence within tolerance $\varepsilon$

---

## IV. CENTRALIZED POWER CONTROL

A centralized power control solution is developed next, which will also serve as a benchmark for Algorithm 1.

### A. Problem formulation

The centralized problem is formulated as the minimization of a global network cost function with respect to the power vector $\mathbf{p} = [p_1, p_2, \ldots, p_K]^T$. Following a scalarization approach [33], a network-wide cost function can be defined by a weighted sum of a global delay cost $D$ and a global energy cost $F$, namely:

$$c(\mathbf{p}) = \rho D(\mathbf{p}) + F(\mathbf{p}) \qquad (22)$$

---

[5]This latter approach would significantly change the problem formulation, leading to different resource allocation problems that are not considered in this work.

where $\rho > 0$ is the weighting coefficient in [Joule/s]. As for $F(\mathbf{p})$ and $D(\mathbf{p})$, both should combine the individual cost functions $\{F_k\}_k$ and $\{D_k\}_k$. Let's start with $F(\mathbf{p})$. Although different combinations can be considered, we focus on the network energy cost-benefit ratio, defined as the ratio between the network total energy consumption and the amount of data reliably transmitted per Joule of consumed energy. This yields:

$$F(\mathbf{p}) = \frac{\sum\limits_{k=1}^{K} p_k + P_{C,k}}{R \sum\limits_{k=1}^{K} S_k(\gamma_k)}. \qquad (23)$$

The metric in (23) can also be seen as the inverse of the popular global energy efficiency metric [31], which represents the energy benefit-cost ratio of a communication network. Thus, (23) represents the network energy cost-benefit ratio, which is a less investigated energy-efficient metric.

As for the delay cost, two cases will be considered. In the first case, we compute $D(\mathbf{p})$ as

$$D(\mathbf{p}) = \frac{1}{K} \sum_{k=1}^{K} D_k = \frac{1}{KR} \sum_{k=1}^{K} \frac{1}{S_k(\gamma_k) - \lambda_k} \qquad (24)$$

which represents the average of the individual delays incurred by the different links of the network. Notice that (24) does not account for the total communication delay incurred by the network, because in practice UEs do not transmit one after the other, but rather in parallel. To account for this, we also consider the case in which

$$D(\mathbf{p}) = \max_k D_k = \frac{1}{R} \frac{1}{\min\limits_k \{S_k(\gamma_k) - \lambda_k\}} \qquad (25)$$

which represents the maximum communication delay incurred in the whole network when one packet is transmitted over all links. By plugging (23)–(25) into (22) leads to the two following global cost functions:

$$c_{\text{sum}}(\mathbf{p}) = \frac{1}{KR} \sum_{k=1}^{K} \frac{\rho}{S_k(\gamma_k) - \lambda_k} + \frac{\sum\limits_{k=1}^{K} p_k + P_{C,k}}{R \sum\limits_{k=1}^{K} S_k(\gamma_k)} \qquad (26)$$

and

$$c_{\min}(\mathbf{p}) = \frac{\rho}{R \min\limits_k \{S_k(\gamma_k) - \lambda_k\}} + \frac{\sum\limits_{k=1}^{K} p_k + P_{C,k}}{R \sum\limits_{k=1}^{K} S_k(\gamma_k)} \qquad (27)$$

that have not been considered in the literature so far. Accordingly, the centralized power control problem is formulated as

$$\arg\min_{\mathbf{p}} \quad c(\mathbf{p}) \qquad (28a)$$

$$\text{subject to} \quad \mathbf{p} \in \mathcal{A} = \mathcal{A}_1 \times \cdots \times \mathcal{A}_K \qquad (28b)$$

wherein the objective $c(\mathbf{p})$ can be taken as either (26) or (27). The first challenge posed by (28) is that it may not be feasible due to the minimum QoS constraints $\{S_k(\gamma_k) \geq \theta_k\}_k$. Aiming mainly at developing a benchmark algorithm, in what follows we relax such constraints and focus on solving

$$\arg\min_{\mathbf{p}} \quad c(\mathbf{p}) \tag{29a}$$

$$\text{subject to} \quad 0 \leq p_k \leq P_{\max,k} \quad \forall k \tag{29b}$$

which is always feasible. Moreover, the maximum value of (29a) subject to (29b) is a lower bound to the optimal solution of (28), since the feasible set of (28) is included in the feasible set of (29).

Even upon relaxing the QoS constraint, solving (29) is still a challenge. Indeed, the objective is a fractional-based function which is neither jointly convex nor jointly pseudo-convex in $\mathbf{p}$. As a result, traditional convex optimization theory and fractional programming approaches can not be used to solve (29) with affordable complexity. To overcome this issue, the framework of *maximum block improvement* (MBI) optimization is used to tackle (29), [26], [43]. In short, the MBI method is an extension of the most popular block coordinate descent method (also known as alternating optimization) [44]. Similar to the the block coordinate descent method, the MBI operates in an iterative manner and partitions the optimization variables into two or more blocks of variables, which are optimized one at a time. Unlike the block coordinate descent method, in each iteration only the variable block which yields the maximum decrement of the objective function is updated. This guarantees that the MBI enjoys the same optimality properties as the block coordinate descent method (i.e., monotonic improvement of the cost function and first-order optimality upon convergence), but under milder assumptions, which are reviewed in Appendix D. In what follows, we first consider the case in which $c(\mathbf{p})$ is given by (26), and then extend the analysis to $c(\mathbf{p})$ given by (27). In both cases, $\mathbf{p}$ will be partitioned into the $K$ blocks $p_1, \ldots, p_K$ such that one transmit power is optimized at each iteration.

### B. Maximum Block Improvement for solving (26)

Applying the MBI algorithm to (29) with $c(\mathbf{p})$ given by (26) amounts to solving for any $k$

$$\arg\min_{p_k} \quad c_{\text{sum}}(p_k, \mathbf{p}_{-k}) \tag{30a}$$

$$\text{subject to} \quad 0 \leq p_k \leq P_{\max,k}. \tag{30b}$$

Notice that, as a function of $p_k$ only, (30a) is the sum of a convex function plus a pseudo-convex function, which is neither convex nor pseudo-convex [31]. This makes (30) not convex. To proceed, let us define

$$\mathcal{S}_k = \left\{ 0 < \bar{p}_k < P_{\max,k} : \frac{\partial c_{\text{sum}}(\bar{p}_k)}{\partial p_k} = 0 \right\} \tag{31}$$

as the set of all feasible stationary points of (30) which can be determined by finding the solutions of the equation $\frac{\partial c_{\text{sum}}}{\partial p_k} = 0$ in the interval $(0, P_{\max,k})$, which is written as

$$-\sum_{\ell=1}^{K} \frac{\rho \frac{\partial S_\ell}{\partial p_k}}{(S_\ell - \lambda_\ell)^2} + \frac{\sum_{\ell=1}^{K} S_\ell(\gamma_\ell) - \left(\sum_{\ell=1}^{K} \frac{\partial S_\ell}{\partial p_k}\right)\left(\sum_{\ell=1}^{K} p_\ell + P_{c,\ell}\right)}{\left(\sum_{\ell=1}^{K} S_\ell(\gamma_\ell)\right)^2} = 0. \tag{32}$$

---

**Algorithm 2** An MBI algorithm to solve (33)

Set $\varepsilon > 0$; $n = 0$; $\mathbf{p}^{(0)}$ any feasible power vector;
**repeat**
    **for** $k = 1$ to $K$ **do**
        $p_k^\star = \arg\min_{p_k \in \mathcal{P}_k} c_{\text{sum}}(p_k)$;
    **end for**
    $n = n+1$; $\bar{k} = \arg\min_k c_{\text{sum}}(p_k^\star, \mathbf{p}^{(n-1)})$;
    $p_{\bar{k}}^{(n)} = p_{\bar{k}}^\star$; $p_k^{(n)} = p_k^{(n-1)}$ for all $k \neq \bar{k}$;
**until** convergence within tolerance $\varepsilon$

---

Then, the following result holds.

*Lemma 3:* The solution to (30) is obtained as:

$$p_k^* = \arg\min_{p_k \in \mathcal{P}_k} c_{\text{sum}}(p_k) \tag{33}$$

wherein $\mathcal{P}_k = \mathcal{S}_k \cup \{0\} \cup \{P_{\max,k}\}$.

*Proof:* The problem in (30) has a differentiable function of one scalar variable as objective, and a compact interval of the real line as feasible set. Thus, by Weirstrass theorem a solution exists. Moreover, it must either lie in the interior of the feasible set, i.e. the open interval $(0, P_{\max,k})$, or on the boundary of the feasible set. But if the solution lies in the open interval $(0, P_{\max,k})$, then (30b) is inactive, thus implying that the solution must be a stationary point of (30a). Hence, the solution is either $p_k = 0$, or $p_k = P_{\max,k}$, or it must belong to $\mathcal{S}_k$. ∎

Finally, on the basis of the results reviewed in Appendix D, and in particular on [26, Theorem 3.1, Corollary 3.2], an MBI algorithm to tackle (30) can be formulated as in Algorithm 2 and the following result holds.

*Proposition 3:* At each iteration of Algorithm 2, (29a) does not increase. Also, every limit point of the sequence generated by Algorithm 2 fulfills the Karush Kuhn Tucker (KKT) first-order optimality conditions of (29).

It is useful to observe that the computational complexity of Algorithm 2 is mainly due to solving the non-linear equation (32) for all $k$ in each iteration of the algorithm, which can be accomplished by standard numerical methods, such as Newton's procedure or bisection methods.

### C. Maximum Block Improvement for Solving (27)

A similar approach can be employed to maximize (27). Unlike (26), $c_{\min}(\mathbf{p})$ is not differentiable due to the min function, which prevents from considering the stationary points of the individual subproblems. To circumvent this issue, a preliminary step is to remove the non-differentiability of $c_{\min}(\mathbf{p})$ by introducing the auxiliary variable $t$ and by reformulating the problem as:

$$\arg\min_{(\mathbf{p},t)} \quad \frac{\rho}{t} + \frac{\sum_{\ell=1}^{K} p_\ell + P_{C,\ell}}{R\sum_{\ell=1}^{K} S_\ell(\gamma_\ell(\mathbf{p}))} \tag{34a}$$

$$\text{subject to} \quad 0 \leq p_\ell \leq P_{\max,\ell} \;\forall \ell \tag{34b}$$

$$t \leq R\big(S_\ell(\gamma_\ell(\mathbf{p})) - \lambda_\ell\big) \;\forall \ell. \tag{34c}$$

The above problem is in the same form of (29) with $c(\mathbf{p}) = c_{\min}^\circ(\mathbf{p})$ and contains only differentiable functions. However, it also requires the optimization of the additional variable $t$. Thus, it is natural to consider the variable blocks $p_k$ for all $k$, plus the additional variable block $t$.

*1) Optimization of $p_k$:* The optimization of $p_k$ for a given $t$ can be carried out as done in Section IV-B by solving:

$$\arg\min_{p_k} \quad c_{\min}^\circ(\mathbf{p}) \tag{35a}$$
$$\text{subject to} \quad 0 \le p_k \le P_{\max,k} \tag{35b}$$
$$RS_\ell(\gamma_\ell(\mathbf{p})) \ge t + \lambda_\ell \ \forall\ \ell \tag{35c}$$

where

$$c_{\min}^\circ(\mathbf{p}) = \frac{\sum_{\ell=1}^K p_\ell + P_{C,\ell}}{\sum_{\ell=1}^K S_\ell(\gamma_\ell(\mathbf{p}))} \tag{36}$$

is obtained from (34b) neglecting the term $\rho/t$, since $t$ and $\rho$ are both fixed and given. It is important to observe that, while among the $K$ constraints in (34b) only one depends on $p_k$ and thus needs to be considered in (35b), all the $K$ constraints in (34c) depend on $p_k$ through the SINR $\gamma_k$, and so they all need to be considered in (35c). Let us denote by $S_\ell^{-1}(\cdot)$ the inverse function of $S_\ell$, and define

$$b_\ell = S_\ell^{-1}\left(\frac{t + \lambda_\ell}{R}\right) \quad \ell = 1,\ldots,K \tag{37}$$
$$\psi_{\ell,k} = \sigma_\ell^2 + \phi_\ell p_\ell + \sum_{j\ne\ell, j\ne k} p_j \beta_{\ell,j} \quad \ell \ne k \tag{38}$$

If $\ell = k$, then (35c) becomes

$$p_k \ge \frac{b_k \omega_k}{\alpha_k - b_k \phi_k} \triangleq P_{\min,k} \tag{39}$$

otherwise, if $\ell \ne k$ we have that

$$p_k \le \frac{\alpha_\ell p_\ell}{b_\ell \beta_{\ell,k}} - \frac{b_\ell \psi_{\ell,k}}{\beta_{\ell,k}} \triangleq \bar{P}_{\ell,k}. \tag{40}$$

Using (39) and (40), and recalling (35b), the feasible set of (35) can be written, for any $k$, as the compact interval $\mathcal{F}_k = [P_{\min,k}, \min_\ell \{P_{\max,k}, \bar{P}_{\ell,k}\}]$. At this point, following the same reasoning of Section IV-B, the solution to (35) will either be a stationary point belonging to the interior of the feasible set, or it will lie on the boundary of the feasible set. Otherwise stated, defining the set of the stationary points of $c_{\min}^\circ(p_k)$ that belong to the interior of the feasible set, namely

$$\mathcal{S}_k = \left\{\bar{p}_k \in \left(P_{\min,k}, \min_\ell\{P_{\max,k}, \bar{P}_{\ell,k}\}\right): \frac{\partial c_{\min}^\circ(\bar{p}_k)}{\partial p_k} = 0\right\} \tag{41}$$

the solution to (35) can be compactly written as

$$p_k^\star = \arg\min_{p_k \in \mathcal{P}_k} c_{\min}^\circ(p_k) \tag{42}$$

where we have defined $\mathcal{P}_k = \mathcal{S}_k \cup \{P_{\min,k}\} \cup \{P_{\max,k}\}$. As for determining the elements of the set $\mathcal{S}_k$, it can be accomplished by finding the solution to the equation $\frac{\partial c_{\min}^\circ}{\partial p_k} = 0$ in the interval $(P_{\min,k}, \min_\ell\{P_{\max,k}, \bar{P}_{\ell,k}\})$, which reads

$$\frac{\sum_{\ell=1}^K S_\ell(\mathbf{p}) - \left(\sum_{\ell=1}^K \frac{\partial S_\ell}{\partial p_k}\right)\left(\sum_{\ell=1}^K p_\ell + P_{C,\ell}\right)}{\left(\sum_{\ell=1}^K S_\ell(\mathbf{p})\right)^2} = 0. \tag{43}$$

*2) Optimization of $t$:* For a given $\mathbf{p}$, the optimization with respect to the auxiliary parameter $t$ is stated as:

$$\arg\min_t \quad \frac{\rho}{t} \tag{44a}$$
$$\text{subject to} \quad t \le R\big(S(\gamma_\ell) - \lambda_\ell\big) \ \forall\ell. \tag{44b}$$

The following result holds.

*Lemma 4:* For any fixed power vector $\mathbf{p}$, the $t$ that solves (44) is given by

$$t^\star = \min_{\ell=1,\ldots,K} \{S(\gamma_\ell) - \lambda_\ell\}. \tag{45}$$

*Proof:* To begin with, let us notice that $S(\gamma_\ell) - \lambda_\ell$ does not depend on $t$ for any $\ell$. Thus, if the solution were some $\bar{t} < t^\star$, then it would always exist a $\tilde{t}$ such that $\bar{t} < \tilde{t} \le t^\star$. It can be readily seen that $\tilde{t}$ is feasible for (44b) and yields a lower value of (44a) than $\bar{t}$, thus contradicting the statement that $\bar{t}$ is the solution. ∎

Finally, based on (42) and (45), an MBI algorithm can be eventually devised along the same lines as Algorithm 2.

## V. NUMERICAL RESULTS

Numerical results are now given for a multicell massive MIMO system with $L = 4$ cells, and 8 users per-cell, for a total of $K = 32$ users. Each cell is a square with edge $500\,\text{m}$ which is served by a base station (BS) with $N = 64$ antennas. In each cell the users are randomly distributed, with a minimum distance of $10\,\text{m}$ from the service BS. All UEs have the same maximum feasible power $P_{\max,k} = P_{\max}$ and hardware-dissipated power $P_{C,k} = P_C = 10\,\text{dBm}$. The receive noise power is $\sigma^2 = FB\mathcal{N}_0$, wherein $F = 3\,\text{dB}$ is the receive noise figure, $B = 180\,\text{kHz}$ is the communication bandwidth, and $\mathcal{N}_0 = -174\,\text{dBm/Hz}$ is the noise spectral density at the receiver. All channels are generated according to the Rayleigh fading model with path loss model as in [45], with power path-loss equal to 3.5. Both hardware impairments at the UEs, and channel estimation errors at the BSs are assumed and modeled following [19], with channel estimation accuracy factor $\tau = 0.3$ and hardware impairment factor $\epsilon = 0.1$. It was shown in [19] that such a scenario leads to an SINR expression which is formally equal to (1), for particular expressions of the coefficients $\{\alpha_k\}_k$, $\{\phi_k\}_k$, $\{\beta_{k,j}\}_{k,j}$, $\{\sigma_k^2\}_k$. For all $k = 1,\ldots,K$, the weight factor has been set to $\rho_k = \rho = 1\,\text{J/s}$, while the adopted efficiency function was (4), with the communication rate $R = 1\,\text{Mbit/s}$, and $\delta_k$ chosen so as to minimize the mean squared error of $S_k(\gamma_k)$ with respect to the packet error probability assuming independent transmissions with a QPSK modulation.

Fig. 1 compares the value of the network-wide cost function (26), versus $P_{\max}$, for the following schemes: (*a*) Algorithm 1, with $\theta_k = \theta = 1 - 10^{-3}$ for all $k$. In case one BR is unfeasible, we relax the QoS constraints to $\theta = 0$; (*b*) Algorithm 1, with $\theta_k = \theta = 1 - 10^{-2}$ for all $k$. In case one BR is unfeasible, we relax the QoS constraints to $\theta = 0$; (c) Algorithm 2 for the minimization of (26). As expected, Scheme (b) performs better than Scheme (a) since enforcing stricter QoS constraints results in a smaller feasible set. In particular, for low values of $P_{\max}$, both schemes perform similarly, because in this range the QoS constraints are not feasible and therefore are relaxed, falling back to the unconstrained case. Instead, for larger values of $P_{\max}$ the QoS are fulfilled by increasing the transmit powers and this increases the cost function, which would otherwise saturate. Specifically, at the equilibrium the QoS constraints will be active for some UEs, which will transmit with the minimum power necessary to fulfill their QoS constraint. Instead, the remaining UEs will have a strictly higher throughput than their minimum requirement, and will transmit with a power that achieves the minimum of their individual cost function in (6). However, this increase of cost function enables to guarantee a lower packet error probability. Finally, it is seen how the centralized benchmark outperforms both non-cooperative schemes, due to the fact that its cooperative nature enables it to suitably manage the multiuser interference.

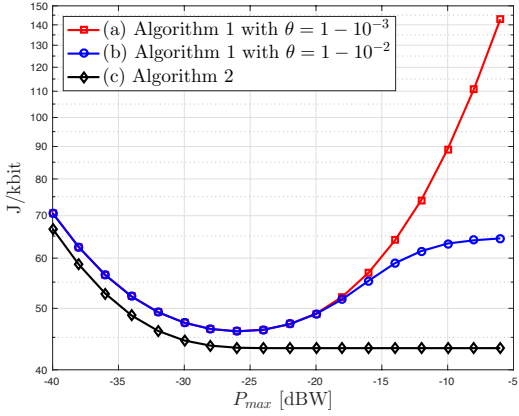

Fig. 1.  $K = 32; N = 64; \epsilon = 10^{-1}; \tau = 0.3$. Average cost (26) versus $P_{\max}$ for: (a) Algorithm 1 with $\theta = 1 - 10^{-3}$; (b) Algorithm 1 with $\theta = 1 - 10^{-2}$; (c) Algorithm 2.

While Fig. 1 considers the overall cost function (26), Fig. 2 shows the achieved values of the energy cost (23) and of the delay cost (24). The two upper plots consider a weighting factor of $\rho = 1\,\text{J/s}$, while the two lower plots consider $\rho = 10\,\text{J/s}$, thus giving more importance to the delay cost. The results show that Scheme (c) achieves the lowest energy cost, followed by Schemes (b) and (a). This is explained recalling that the stricter the QoS constraint, the larger the transmit power required to fulfill it. Instead, the different schemes perform similarly in terms of delay cost, which is explained recalling the expression of the delay cost in (24), and that for both schemes it holds $1 - 10^{-2} \leq S_k(\gamma_k) \leq 1$. Thus, the difference in the values of $S_k$ achieved by the two schemes is negligible. Next, it is also seen that Scheme (c) has a slightly larger delay cost than Schemes (a) and (b) when $\rho = 1\,\text{J/s}$, whereas this does not happen when $\rho = 10\,\text{J/s}$. To explain this seemingly counterintuitive result, we observe that Scheme (c) is based on Algorithm 2, which minimizes the cost function (26), given by the sum of the energy cost (23) and of the delay cost (24). So, Algorithm 2 will yield a lower value of (26), but this does not imply that it will also yield a lower value of both summands (23) and (24). In particular, when $\rho_k = \rho = 1\,\text{J/s}$, the predominant term in (26) turns out to be the energy cost (23). Instead, the scenario changes when $\rho_k = \rho = 10\,\text{J/s}$, since this enhances the priority of the delay cost in the optimization process.

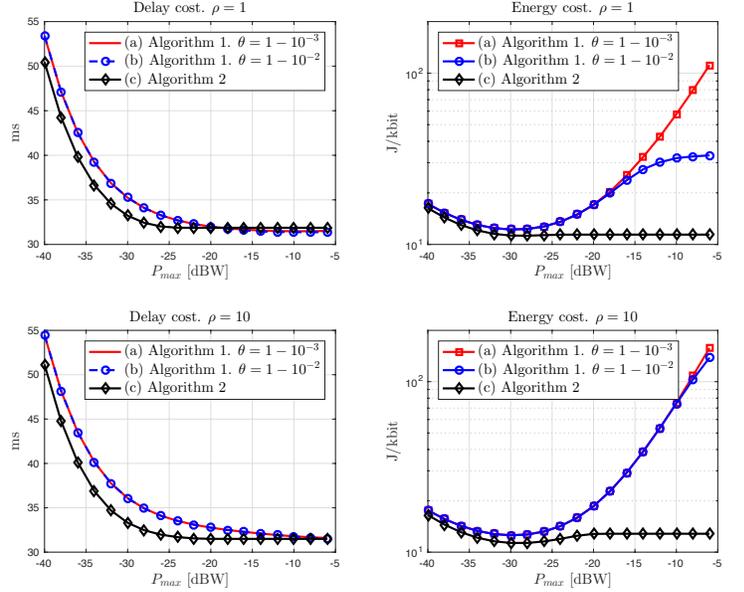

Fig. 2.  $K = 32; N = 64; \epsilon = 10^{-1}; \tau = 0.3$; Upper plots: $\rho = 1\,\text{J/s}$; Lower plots: $\rho = 10\,\text{J/s}$. Average energy cost (23) and delay cost (24) versus $P_{\max}$ for: (a) Algorithm 1 with $\theta = 1 - 10^{-3}$; (b) Algorithm 1 with $\theta = 1 - 10^{-2}$; (c) Algorithm 2.

Unlike previous figures that focused on average costs, Fig. 3 reports the minimum and maximum values among the users cost functions $c_k$ obtained upon convergence of the non-cooperative Algorithm 1, considering both $\theta = 1 - 10^{-2}$ and $\theta = 1 - 10^{-3}$. It is seen that as the transmit power $P_{\max}$ increases, all users tend to have the same cost function at the equilibrium. This means that, if enough transmit power is available, proper interference management is possible, and the game Nash equilibrium corresponds to a fair scenario, in which all users enjoy similar performance.

Fig. 4 shows the average UEs' transmit power for Schemes (a), (b), and (c), with $\rho_k = \rho = 1\,\text{J/s}$. The results show that Scheme (a) requires a larger transmit power than Scheme (b), as a consequence of the stricter QoS constraint to be fulfilled. Moreover, Scheme (c) requires the lowest transmit power consumption, due to the fact that no QoS constraints are enforced in Algorithm 2.

Fig. 5 compares the minimization of the two centralized metrics (26) and (27) by Algorithm 2. The upper plot reports the value of (26) obtained by the power allocation resulting from the minimization of (26) and of (27). The lower plot shows the values of (27) obtained by the minimization of the two metrics. It is seen that minimizing either (26) or (27) leads to similar results in terms of both metrics, especially for larger $P_{\max}$, which allows for a better interference management.

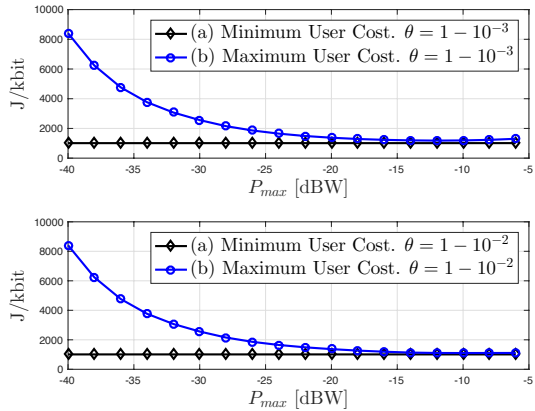

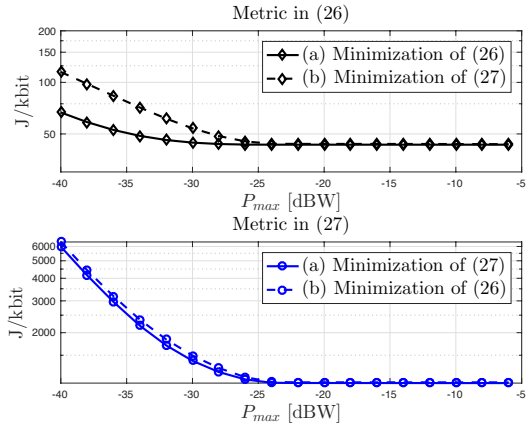

Fig. 3. $K = 32; N = 64; \epsilon = 10^{-1}; \tau = 0.3; \rho = 1\,\text{J/s}$. Upper plot: (*a*) Minimum users' cost function versus $P_{\max}$ by Algorithm 1 with $\theta = 1 - 10^{-3}$; (*b*) Maximum users' cost function versus $P_{\max}$ by Algorithm 1 with $\theta = 1 - 10^{-3}$; Lower plot: (*a*) Minimum users' cost function versus $P_{\max}$ by Algorithm 1 with $\theta = 1 - 10^{-2}$; (*b*) Maximum users' cost function versus $P_{\max}$ by Algorithm 1 with $\theta = 1 - 10^{-2}$.

Fig. 5. $K = 32; N = 64; \epsilon = 10^{-1}; \tau = 0.3; \rho = 1\,\text{J/s}$. Upper plot: Average cost (26) versus $P_{\max}$ for: (*a*) Minimization of (26) by Algorithm 2; (*b*) Minimization of (27) by Algorithm 2. Lower plot: Average cost (27) versus $P_{\max}$ for: (*a*) Minimization of (27) by Algorithm 2; (*b*) Minimization of (26) by Algorithm 2.

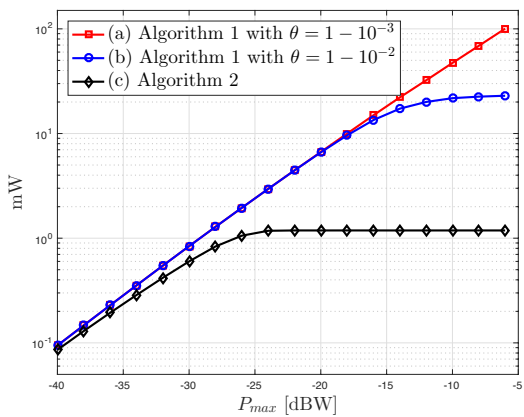

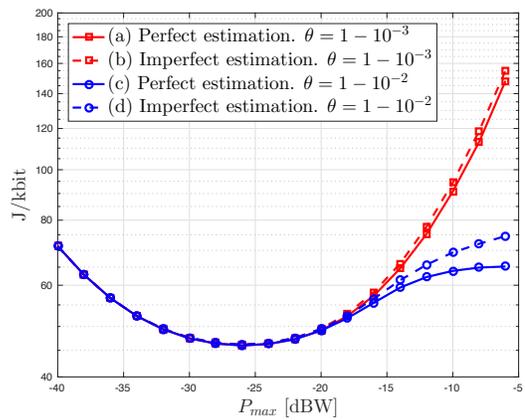

Fig. 4. $K = 32; N = 64; \epsilon = 10^{-1}; \tau = 0.3; \rho = 1\,\text{J/s}$. Average transmit power versus $P_{\max}$ for: (*a*) Algorithm 1 with $\theta = 1 - 10^{-3}$; (*b*) Algorithm 1 with $\theta = 1 - 10^{-2}$; (*c*) Algorithm 2.

Fig. 6. $K = 32; N = 64; \epsilon = 10^{-1}; \tau = 0.3; \rho = 1\,\text{J/s}$. Average cost (26) versus $P_{\max}$ for: (*a*) Algorithm 1 with $\theta = 1 - 10^{-3}$ and perfect parameter estimation; (*b*) Algorithm 1 with $\theta = 1 - 10^{-3}$ and imperfect parameter estimation; (*c*) Algorithm 1 with $\theta = 1 - 10^{-2}$ and perfect parameter estimation; (*d*) Algorithm 1 with $\theta = 1 - 10^{-2}$ and imperfect parameter estimation.

Fig. 6 compares the average cost (26) by the power allocation upon convergence of Algorithm 1 implemented by: 1) using the true values of $\{\alpha_k, \phi_k, \gamma_k\}$; 2) using estimated values $\{\hat{\alpha}_k, \hat{\phi}_k, \hat{\gamma}_k\}$ equal to the true values plus a perturbation term whose magnitude has been randomly generated between zero and 30% of the true value. However, when computing the achieved value of (26) for the two different power allocations, the true parameters $\{\alpha_k, \phi_k\}$ (and also the true SINRs up to the different power allocation) have been used. The results show that Algorithm 1 is quite robust to parameters perturbations, especially for low transmit powers and more demanding QoS constraints. Indeed, in these situations, at the equilibrium every user tends to transmit with full power, regardless of the particular set of parameters that is used.

Finally, Table I reports the average number of iterations required for Algorithm 1 to converge, where convergence is declared when $\|\mathbf{p}^{(n)} - \mathbf{p}^{(n-1)}\|^2/\|\mathbf{p}^{(n)}\|^2 \leq 10^{-4}$, with $\mathbf{p}^{(n)}$ the vector of the players' powers after iteration $n$. Convergence occurs after a limited number of iterations, which tends to increase for larger $P_{\max}$, because increasing $P_{\max}$ results in a larger feasible set. This shows that the proposed non-cooperative approach has a limited computational complexity, thereby lending itself to a simple implementation in practical systems. It is also seen that the number of iterations is higher when stricter QoS constraints are enforced, because in this case the users tend to increase their transmit powers, creating more significant interference, which in turn enhances the coupling among the best-responses and requires more best-response rounds to eventually find an equilibrium.

## VI. CONCLUSIONS

This work proposed a new approach for power control in wireless networks, jointly optimizing both the energy efficiency and the delay of the communication, subject to QoS constraints. The problem was formulated as a non-cooperative generalized game, deriving conditions for the feasibility of the players' best-responses, for the existence of a unique generalized NE, and for the convergence of the BRD. Leveraging these results, a provably convergent, low-complexity, and fully decentralized power control algorithm was developed, and its performance benchmarked against a centralized scheme based on the MBI method. The results indicated a relatively limited gap between the distributed and centralized algorithms, and that the presence of QoS constraints enables a significant performance improvement at the expense of a moderate increase of energy consumption.

TABLE I
$K = 32; N = 64; \epsilon = 10^{-1}; \tau = 0.3$. ITERATION NUMBER OF
ALGORITHM 1 VERSUS $P_{\max}$. $\theta = 1 - 10^{-3}; \theta = 1 - 10^{-2}$.

| $P_{\max}$ | $\theta = 1 - 10^{-2}$ | $\theta = 1 - 10^{-3}$ |
|---|---|---|
| $-40$ dBW | 2.31 | 2.31 |
| $-30$ dBW | 3.56 | 3.56 |
| $-20$ dBW | 5.42 | 9.23 |
| $-10$ dBW | 12.36 | 35.11 |

## APPENDIX A
## PROOF OF PROPOSITION 1

The stationarity condition $\frac{\partial c_k}{\partial p_k} = 0$ yields the equation

$$\frac{S_k(\gamma_k)}{S'_k(\gamma_k)\frac{\partial \gamma_k}{\partial p_k}} - p_k - \frac{\rho_k S_k^2(\gamma_k)}{(S_k(\gamma_k) - \lambda_k)^2} = P_{C,k}. \quad (46)$$

Let us define the positive power $p_{\lambda_k}$ as the power level such that $S_k(\gamma_k) = \lambda_k$. Since $\theta_k > \lambda_k$, it holds $p_{\lambda_k} < P_{\min,k}$. Then, let us show that (46) has a unique, solution in the set $p_k > p_{\lambda_k}$. To begin with, computing the derivative of the left-hand side (LHS) of (46) yields

$$-S_k(\gamma_k)\frac{\frac{\partial^2 \gamma_k}{\partial p_k^2}S'_k(\gamma_k) + S''_k(\gamma_k)\left(\frac{\partial \gamma_k}{\partial p_k}\right)^2}{\left(\frac{\partial \gamma_k}{\partial p_k}S'_k(\gamma_k)\right)^2}$$
$$+ \frac{2\tilde{\rho}_k \lambda_k S_k(\gamma_k) S'_k(\gamma_k) \frac{\partial \gamma_k}{\partial p_k}}{(S_k(\gamma_k) - \lambda_k)^3} \quad (47)$$

which is a positive quantity due to facts that $S_k(\gamma_k)$ and $\gamma_k(p_k)$ are both non-negative, strictly increasing, and strictly concave functions. So, the LHS of (46) is a strictly increasing function of $p_k$. Moreover, for $p_k \to p_{\lambda_k}$, the LHS of (46) tends to $-\infty$. So, (46) has a unique solution if its LHS tends to $+\infty$ for $p_k \to +\infty$. This is equivalent to showing that the derivative of the LHS of (46) does not tend to 0 for $p_k \to +\infty$. This can be shown by plugging the expression of $\gamma_k$ in (1), into (47), which yields

$$\frac{2\phi_k S_k(\gamma_k)}{\alpha_k \omega_k S'_k(\gamma_k)}(\omega_k + \phi_k p_k) - \frac{S_k(\gamma_k) S''_k(\gamma_k)}{(S'_k(\gamma_k))^2}$$
$$+ \frac{2\tilde{\rho}_k \lambda_k S_k(\gamma_k) S'_k(\gamma_k) \alpha_k \omega_k}{(\omega_k + \phi_k p_k)(S_k(\gamma_k) - \lambda_k)^3}. \quad (48)$$

The first term in (48) can be seen to be diverging for $p_k \to \infty$, by virtue of Properties 1-4. Since the remaining terms in (48) are non-negative (again due to the concavity of $S_k$), we conclude that (48) diverges, and this ultimately implies that (46) has a unique, solution in the set $p_k > p_{\lambda,k}$.

In a similar way it can be shown that the objective (9a) is decreasing for $p_{k,\lambda} < p_k \leq \bar{p}_k$ and increasing for $p_k > \bar{p}_k$, which finally implies the thesis.

## APPENDIX B
## PROOF OF LEMMA 2

Let us denote by $g_k(p_k, \omega_k)$ the LHS of (46). Then, $\frac{\partial g_k}{\partial \omega_k}$ is given by

$$\frac{\left((S'_k(\gamma_k))^2 - S_k(\gamma_k)S''_k(\gamma_k)\right)\frac{\partial \gamma_k}{\partial p_k}\frac{\partial \gamma_k}{\partial \omega_k} - S_k(\gamma_k)S'_k(\gamma_k)\frac{\partial^2 \gamma_k}{\partial p_k \partial \omega_k}}{\left(\frac{\partial \gamma_k}{\partial p_k}S'_k(\gamma_k)\right)^2}$$
$$+ \frac{2\tilde{\rho}_k \lambda_k S_k(\gamma_k) S'_k(\gamma_k) \frac{\partial \gamma_k}{\partial \omega_k}}{(S_k(\gamma_k) - \lambda_k)^3}$$
(49)

Given Properties 1-4 of $S_k(\gamma_k)$, and observing that $\gamma_k$ is decreasing with respect to $\omega_k$, it follows that the first term in (49) is negative. As for the second term, plugging in the expression for the derivatives of $\gamma_k$ we obtain that it is non-positive if

$$(S'_k(\gamma_k))^2 - S_k(\gamma_k)S''_k(\gamma_k) + \frac{S_k(\gamma_k)S'_k(\gamma_k)}{\gamma_k}\frac{\phi_k p_k - \omega_k}{\omega_k} \geq 0. \quad (50)$$

Since the function $(\phi_k p_k - \omega_k)/\omega_k \geq -1$, it follows that in order for (50) to hold for all $\omega_k > 0$, it must be true that $(\phi_k p_k - \omega_k)/\omega_k = -1$. Then, (50) becomes

$$-(S'_k(\gamma_k))^2 + S_k(\gamma_k)S''_k(\gamma_k) + \frac{S_k(\gamma_k)S'_k(\gamma_k)}{\gamma_k} \leq 0 \quad (51)$$

which is equivalent to (19).

## APPENDIX C
## GENERALIZED CONCAVITY AND FRACTIONAL PROGRAMMING

We limit our review to basic results. For a more detailed review, we refer the reader to [31] and [46, Chapters 3, 4].

*Definition 1 (Pseudo-convexity):* Let $\mathcal{C} \subseteq \mathbb{R}^n$ be a convex set. Then $r : \mathcal{C} \to \mathbb{R}$ is pseudo-convex if and only if, $\forall \ \mathbf{x}_1, \mathbf{x}_2 \in \mathcal{C}$, it is differentiable and $r(\mathbf{x}_2) < r(\mathbf{x}_1) \Rightarrow \nabla (r(\mathbf{x}_1))^T (\mathbf{x}_2 - \mathbf{x}_1) < 0$.

In a similar way we can define strict pseudo-convexity.

*Definition 2 (Strict pseudo-convexity):* Let $\mathcal{C} \subseteq \mathbb{R}^n$ be a convex set. Then $r : \mathcal{C} \to \mathbb{R}$ is strictly pseudo-concave (PC) if and only if, for all $\mathbf{x}_1 \neq \mathbf{x}_2 \in \mathcal{C}$, it is differentiable and

$$r(\mathbf{x}_2) \leq r(\mathbf{x}_1) \Rightarrow \nabla (r(\mathbf{x}_1))^T (\mathbf{x}_2 - \mathbf{x}_1) < 0. \quad (52)$$

The interest for PC functions stems from the following result.

*Proposition 4:* Let $r : \mathcal{C} \to \mathbb{R}$ be a pseudo-convex function. Then,

(a) If $\mathbf{x}^\star$ is a stationary point for $r$, then it is a global minimizer for $r$;

(b) The KKT conditions for the problem of minimizing $r$ subject to convex constraints are necessary and sufficient conditions for optimality subject to constraint qualification conditions;

(c) If $r$ is strictly pseudo-convex, then a unique minimizer exists.

Pseudo-convexity plays a key-role in the optimization of fractional functions, due to the following result.

*Proposition 5:* Let $r(\mathbf{x}) = \dfrac{f(\mathbf{x})}{g(\mathbf{x})}$, with $f : \mathcal{C} \subseteq \mathbb{R}^n \to \mathbb{R}$ and $g : \mathcal{C} \subseteq \mathbb{R}^n \to \mathbb{R}_+$. If $f$ is non-negative, differentiable, and convex, while $g$ is differentiable and concave, then $r$ is pseudo-convex. If $g$ is affine, the non-negativity of $f$ can be relaxed. Strict pseudo-convexity holds if either $f$ is strictly convex, or $g$ is strictly concave.

*Definition 3 (Fractional program):* Let $\mathcal{C} \subseteq \mathbb{R}^n$ be a convex set, and consider the functions $f : \mathcal{C} \to \mathbb{R}$ and $g : \mathcal{C} \to \mathbb{R}^+$. A fractional program is the optimization problem

$$\min_{\mathbf{x} \in \mathcal{C}} \frac{f(\mathbf{x})}{g(\mathbf{x})}. \tag{53}$$

*Proposition 6:* An $\mathbf{x}^\star \in \mathcal{C}$ solves (53) if and only if $\mathbf{x}^\star = \arg\min_{\mathbf{x} \in \mathcal{C}} \{f(\mathbf{x}) - \lambda^\star g(\mathbf{x})\}$, with $\lambda^\star$ being the unique zero of $F(\lambda) = \min_{\mathbf{x} \in \mathcal{C}} \{f(\mathbf{x}) - \lambda g(\mathbf{x})\}$.

This result allows us to solve (53) by finding the zero of $F(\lambda)$. An efficient algorithm to do so is Dinkelbach's algorithm [47], reported in Algorithm 3 for the reader's convenience. It can be seen that if $f(\mathbf{x})$ and $g(\mathbf{x})$ are convex and concave, respectively, and if $\mathcal{C}$ is a convex set, then Dinkelbach's algorithm requires to solve one convex problem in each iteration.[6] Moreover, the convergence rate of Dinkelbach's algorithm is known to be super-linear [47].

---
**Algorithm 3** Dinkelbach's algorithm
---
Set $\varepsilon > 0$; $\lambda = 0$;
**repeat**
    $\mathbf{x}^\star = \arg\min_{\mathbf{x} \in \mathcal{C}} \{f(\mathbf{x}) - \lambda g(\mathbf{x})\}$
    $F = f(\mathbf{x}^\star) - \lambda g(\mathbf{x}^\star)$; $\lambda = f(\mathbf{x}^\star)/g(\mathbf{x}^\star)$;
**until** $F \leq \varepsilon$

---

It should be mentioned that alternative solution methods are also available for single-ratio fractional programs [31], which are not reviewed here due to space constraints. However, just as Dinkelbach's algorithm, all available methods require $f(\mathbf{x})$ to be convex, $g(\mathbf{x})$ to be concave, and the constraint set to be convex, in order to exhibit affordable complexity.

## APPENDIX D
## MAXIMUM BLOCK IMPROVEMENT

The MBI method is an enhancement of the more popular alternating optimization method. Just like the alternating minimization approach, the MBI is an iterative method which optimizes one variable (block) at a time, while keeping the other variables fixed. However, at the end of each iteration, only one variable block is updated, namely the one yielding the maximum decrement of the objective function. More formally, let us consider the optimization program:

$$\min_{\mathbf{p}} \; g(p_1, \ldots, p_K) \tag{54a}$$
$$\text{s.t.} \; h_i(\mathbf{p}) \leq 0 \; \forall \, i = 1, \ldots, I. \tag{54b}$$

In each iteration, the MBI algorithm solves the $K$ problems

$$\min_{p_k} \; g(p_k, \mathbf{p}_{-k}) \tag{55a}$$
$$\text{s.t.} \; h_i(p_k, \mathbf{p}_{-k}) \leq 0 \; \forall \, i = 1, \ldots, I \tag{55b}$$

for all $k = 1 \ldots, K$, and then updates $p_j$, with $j$ being the index of the variable coordinate which led to the largest decrement of the cost function $g$. The formal procedure is as follows.

---
**Algorithm 4** MBI algorithm
---
Set $\varepsilon > 0$; $n = 0$;
Let $\mathbf{p}^{(n)}$ be any feasible power vector;
**repeat**
    **for** $k = 1$ to $K$ **do**
        Solve (55) and let $q_k$ be the solution;
    **end for**
    $n = n + 1$; $\bar{k} = \arg\min\limits_{1 \leq k \leq K} g(q_k, \mathbf{p}_{-k})$;
    $p_{\bar{k}}^{(n)} = q_{\bar{k}}$; $p_k^{(n)} = p_k^{(n-1)}$ for all $k \neq \bar{k}$;
**until** $\left| g\left(\mathbf{p}^{(n)}\right) - g\left(\mathbf{p}^{(n-1)}\right) \right| \leq \varepsilon$

---

It is clear that by construction the MBI method monotonically reduces the value of the cost function, and thus converges in the value of the objective. Indeed, under the very mild assumption of continuous objective and compact feasible set, (54) must have a finite solution, thus implying that the objective function in (54) is lower-bounded. Moreover, the MBI method enjoys the following additional optimality property, as proved in [26, Theorem 3.1, Corollary 3.2].

*Proposition 7:* Assume the objective function is (54) is differentiable and that the feasible set is compact. Also, assume that the constraint functions are decoupled in the optimization variable blocks. Then, any limit point of the sequence $\{\mathbf{p}_n\}_n$ fulfills the KKT first-order optimality conditions of Problem (54).

*Remark 3:* It is worth observing that this result is an extension of the more popular result from [44] which applies to the block coordinate descent method (also known as alternating optimization). Indeed, the optimality properties stated in Proposition 7 hold for the block coordinate descent method when the additional assumption is made, that the solution to the subproblem (55) is unique for any $k$. Instead, this assumption is not required for the MBI method.

---

[6] It is also required that $\lambda \geq 0$ in each iteration. This can be shown to always hold if the algorithm starts with $\lambda = 0$, and provided $\min_{\mathbf{x}} f(\mathbf{x}) \geq 0$.